\documentclass[12pt]{article}

\usepackage{amsmath,amsfonts}
\usepackage{amssymb}
\usepackage{amsthm}

\newcount\N  \newcount\M  \newdimen\SIZE  \newdimen\INC
\def\YGBOX#1#2#3{
      \N=#1  \M=1  \INC=#2pt  \advance\INC by .#3pt  
      \vbox{
         \loop\ifnum\M>0
            \M=\N
            \divide\N by 10         
            \multiply\N by 10
            \advance\M by -\N
            \divide\N by 10
            \SIZE=\INC
            \multiply\SIZE by \M    
            \advance\SIZE by .#3pt
             \hrule  width \SIZE  height .#3pt
              \hbox{\loop\ifnum\M>0                      
                        \vrule  height #2pt  width .#3pt 
                        \hskip #2pt
                        \advance\M by -1  \repeat
                        \vrule  width .#3pt }
             \hrule  width \SIZE  height .#3pt
            \vskip -.#3pt
         \repeat } }

\newcount\N  \newcount\M  \newdimen\SIZE  \newdimen\INC
\def\YGBOXC#1#2#3{
      \N=#1  \M=1  \INC=#2pt  \advance\INC by .#3pt  
      \vcenter{\vbox{
         \loop\ifnum\M>0
            \M=\N
            \divide\N by 10         
            \multiply\N by 10
            \advance\M by -\N
            \divide\N by 10
            \SIZE=\INC
            \multiply\SIZE by \M    
            \advance\SIZE by .#3pt
             \hrule  width \SIZE  height .#3pt
              \hbox{\loop\ifnum\M>0                      
                        \vrule  height #2pt  width .#3pt 
                        \hskip #2pt
                        \advance\M by -1  \repeat
                        \vrule  width .#3pt }
             \hrule  width \SIZE  height .#3pt
            \vskip -.#3pt
         \repeat } } }

\def\youngc#1{{             
       \mathchoice{\YGBOXC{#1}61}{\YGBOXC{#1}61}{\YGBOXC{#1}41}{\YGBOXC{#1}31}}}

\title{\bf Toda equation  
and special polynomials 
associated with the Garnier system}

\date{}
\author{Teruhisa TSUDA  \\
Department of Mathematics, 
Kobe University, \\
Rokko, Kobe 657-8501,
Japan.
\\
e-mail: tudateru@ms.u-tokyo.ac.jp
}

\begin{document}
\maketitle

\renewcommand{\thefootnote}{\fnsymbol{footnote}}
\footnotetext{This article is based on the results in the author's Ph.D thesis \cite{T6}.} 

\newtheorem{thm}{Theorem}[section]
\newtheorem{prop}[thm]{Proposition}
\newtheorem{cor}[thm]{Corollary}
\newtheorem{lemma}[thm]{Lemma}
\newtheorem*{dfn}{Definition}
\newtheorem*{claim}{Claim}
\newtheorem{remark}[thm]{\it Remark}
\newtheorem{example}[thm]{\it Example}

\numberwithin{equation}{section}

\def\pf{\noindent{\it Proof.\quad}}
\def\qed{\hfill $\blacksquare$}
\def\Remark{\medskip\noindent {\it Remark.}\quad}
\def\rank{{\rm rank}~}
\def\corank{{\rm corank}~}
\def\id{{\rm id}}
\def\dim{{\rm dim}~}
\def\Example{\medskip\noindent {\it Example.}\quad}

\renewcommand{\thefootnote}{\fnsymbol{footnote}}

\begin{abstract}
We prove that 
a certain sequence of tau functions
of the Garnier system
satisfies Toda equation. 
We construct a class of algebraic solutions of the system 
by the use of Toda equation;
then show that the associated tau functions are expressed in terms of the universal character,
which is a generalization of Schur polynomial attached to a pair of partitions.
\end{abstract}

\newpage
\section*{Introduction} 
The {\it Garnier system} is 
the following completely integrable Hamiltonian system of partial differential equations
(see \cite{G,IKSY,KiO}):
\begin{subequations}   \label{subeq:hn}
\begin{equation} \label{eq:gn}
\frac{\partial q_i }{\partial s_j} =\frac{\partial H_j }{\partial p_i},  \quad
\frac{\partial p_i }{\partial s_j} = - \frac{\partial H_j }{\partial q_i}, 
\quad  (i,j = 1, \ldots, N),
\end{equation}
with Hamiltonians
\begin{eqnarray} 
s_i(s_i-1)H_i &=& q_i \left(\alpha +\displaystyle{ \sum_j q_j p_j}\right)\left(\alpha+\kappa_\infty + \displaystyle{ \sum_j q_j p_j} \right)  +s_ip_i(q_ip_i-\theta_i) 
\nonumber 
\\
&& - \sum_{j(\neq i)} R_{ji}(q_jp_j-\theta_j)q_ip_j
   - \sum_{j(\neq i)} S_{ij}(q_ip_i-\theta_i)q_jp_i 
\nonumber \\
&& - \sum_{j(\neq i)} R_{ij}q_jp_j(q_ip_i-\theta_i) 
   - \sum_{j(\neq i)} R_{ij}q_ip_i(q_jp_j-\theta_j) 
\nonumber \\
&&   -(s_i+1)(q_ip_i-\theta_i)q_ip_i + (\kappa_1 s_i+\kappa_0-1)q_ip_i ,
\label{eq:hamiltonianofgn}
\end{eqnarray}
\end{subequations}
where
$R_{ij} = {s_i(s_j-1)}/{(s_j-s_i)}$, 
$S_{ij} =  {s_i(s_i-1)}/{(s_i-s_j)}$
and
\begin{equation}
\alpha = 
-\frac{1}{2}\left(\kappa_0+ \kappa_1 + \kappa_\infty + 
\sum_i \theta_i -1 \right).
\end{equation}
Here the symbols
$\sum_i$ and $\sum_{i (\neq j)}$
stand for 
the summation over $i=1,\ldots,N$
and over $i=1,\ldots,j-1,j+1,\ldots,N$,
respectively.
System (\ref{subeq:hn}) contains $N+3$ constant parameters
\begin{equation}
\vec{\kappa}=(\kappa_0,\kappa_1,\kappa_\infty,\theta_1,\ldots, \theta_N  ) \in {\mathbb C}^{N+3},
\end{equation}
so that  we often denote  it by
${\cal H}_N = {\cal H}_N(\vec{\kappa}) 
=  {\cal H}_N(q,p,s,H;\vec{\kappa})$,
and so on. 
The Garnier system governs the monodromy preserving deformation of 
a Fuchsian differential equation with $N+3$  singularities
and is an extension of the sixth Painlev\'e equation $P_{\rm VI}$;
for $N=1$, (\ref{subeq:hn}) is equivalent to the Hamiltonian system of $P_{\rm VI}$
(see \cite{O}), in fact.

In this paper, 
we prove that a certain sequence of $\tau$-functions of the Garnier system satisfies Toda equation.
We construct a class of algebraic solutions of the system by using Toda equation;
then show that the corresponding $\tau$-functions are expressed
in terms of the universal character,
which is a generalization of Schur polynomial
attached to  a pair of partitions.

First we introduce a group of birational canonical transformations of the Garnier system ${\cal H}_N$.
The group forms an infinite group which contains a translation ${\mathbb Z}$;
see Sect.~\ref{sect:birat}.
We define a function $\tau=\tau(s;\vec{\kappa})$, 
called the {\it $\tau$-function}
(see \cite{IKSY, KiO}),
by 
\begin{equation}
{\rm d} \log \tau=\sum_i H_i {\rm d}s_i.
\end{equation}
By the use of birational symmetries of ${\cal H}_N$, we have the 
\begin{thm} 
A certain sequence of $\tau$-functions 
$\{ \tau_n  | n \in{\mathbb Z}\}$
satisfies the Toda equation{\rm:}
\begin{equation}
X Y \log \tau_n = c_n \frac{ \tau_{n-1} \tau_{n+1}}{\tau_n^2}, 
\end{equation}
where $X$, $Y$ 
being vector fields such that $[X,Y]=0$
and $c_n$ a nonzero constant. 
\end{thm}
\noindent
(See Theorem \ref{thm:toda}.)

Consider the fixed point of a certain birational symmetry, 
we obtain an algebraic solution of the Garnier system.
For example, if $\kappa_0 = \kappa_1 = 1/2$,  
then ${\cal H}_N$ admits an algebraic solution
\begin{equation}
(q_i,p_i)=\left(\frac{\theta_i \sqrt{s_i}}{\kappa_\infty},\frac{\kappa_\infty}{2\sqrt{s_i}}\right), \quad
i=1,\ldots ,N.
\end{equation}
Applying the action of the group of birational symmetries,
we thus have the 
\begin{thm}   \label{thm:intro2}
If two components of the parameter 
$\vec{\kappa}=
(\kappa_0,\kappa_1,\kappa_\infty,\theta_1,\ldots, \theta_N)$
are half integers
then the Garnier system  ${\cal H}_N$ admits an algebraic solution.
\end{thm}
\noindent
(See Theorem~\ref{thm:algsol}.)

Secondly 
we investigate the $\tau$-functions associated with algebraic solutions of the Garnier system.
Starting from the $\tau$-function corresponding to an algebraic solution, we determine
a sequence of $\tau$-functions by means of Toda equation. 
Such a sequence of $\tau$-functions is converted to polynomials
$T_{m,n}=T_{m,n}(t)$ $(m,n \in{\mathbb Z})$
through a certain normalization,
where $t=(t_1,\ldots,t_N)$ and $t_i=\sqrt{s_i}$.
We call $T_{m,n}$ 
{\it special polynomials}
associated with algebraic solutions of ${\cal H}_N$
(see Sect.~\ref{sect:alg}).
Algebraic solutions are explicitly written in terms of the special polynomials.
\begin{thm}  
If 
$\kappa_0 =1/2+ m+n$,
$\kappa_1 = 1/2+ m-n$ $(m,n \in {\mathbb Z})$,
then 
${\cal H}_N$ 
admits an algebraic solution given by
\begin{equation}
\begin{array}{l}
q_i = \frac{ \displaystyle{t_i \frac{\partial}{\partial t_i} \log \frac{T_{m+1,n}}{ T_{m,n+1}} } }
{ \displaystyle \sum_j t_j \frac{\partial}{\partial t_j} \log \frac{T_{m+1,n}}{ T_{m,n+1}} 
-2m+2n-1},
\\ 
\displaystyle
2 q_i  p_i =
\theta_i+m+n + t_i \frac{\partial}{\partial t_i} \log \frac{T_{m,n}}{ T_{m,n+1}} .
\end{array}
\end{equation}
\end{thm}
\noindent
(See Theorem~\ref{thm:algsolinT}.)
Note that  we immediately obtain also the expressions of the other algebraic solutions in Theorem~\ref{thm:intro2}, 
via the birational symmetries of ${\cal H}_N$.
Finally we give an explicit formula for $T_{m,n}$ in terms of 
the universal character
(see \cite{Ko,T3}),
which is a generalization of Schur polynomial.
\begin{thm}  \label{thm:intro4}
The special polynomials $T_{m,n}(t)$
$(m,n \in {\mathbb Z})$ 
is expressed as follows{\rm:}
\begin{equation}
T_{m,n}(t) = N_{m,n}
S_{[\lambda,\mu]}(x,y).
\end{equation}
Here $S_{[\lambda,\mu]}(x,y)=S_{[\lambda,\mu]}(x_1,x_2,\ldots,y_1,y_2,\ldots)$
denotes the universal character attached to a pair of partitions
\begin{equation}
\lambda = (u,u-1, \ldots,2,1),  \quad
\mu=(v,v-1,\ldots,2,1),
\end{equation}
with
$u=|n-m-{1}/{2}|-{1}/{2}$,  
$v=|n+m-{1}/{2}|-{1}/{2}${\rm;}
$N_{m,n}$ is a certain normalization factor,
and 
\begin{equation}
x_n=\frac{-\kappa_\infty+\sum_{i} \theta_i t_i^n}{n}, 
\quad 
y_n=\frac{-\kappa_\infty+\sum_{i}\theta_i t_i^{-n}}{n}.
\end{equation}
\end{thm}
\noindent
(See Theorem \ref{thm:Tmnuc} and also Corollary~\ref{cor:Tmnuc}.) 
Recall that the universal character 
is the irreducible character of 
a rational representation of $GL(n)$,
while Schur polynomial that of a polynomial representation;
see \cite{Ko}.
Hence  Theorem~\ref{thm:intro4} shows us a relationship between 
the representation theory of $GL(n)$
and the Garnier system,
or the theory of monodromy preserving deformation.

 We propose in \cite{T3}
an infinite dimensional integrable system 
characterized by the universal characters,
called the UC hierarchy;
and regard it as an extension of the KP hierarchy.
Since all the universal characters are solutions of 
the UC hierarchy,
it would be an interesting problem to construct
a certain reduction procedure from the hierarchy
to the Garnier system;
{\it cf.} \cite{T5}.
\\

In Sect.~\ref{sect:birat},
we present a group of birational canonical transformations 
of the Garnier system ${\cal H}_N$.
In Sect.~\ref{sect:toda},
we prove that 
a certain sequence of $\tau$-functions 
satisfies Toda equation.
In Sect.~\ref{sect:alg},
we construct a class of algebraic solutions of ${\cal H}_N$
by using Toda equation;
then show that the associated $\tau$-functions
are  explicitly written 
in terms of the universal characters.
Sect.~\ref{sect:proof} is devoted to the verification of Theorem~\ref{thm:Tmnuc}.

\section{Birational symmetry}  \label{sect:birat}
First we introduce a group of birational canonical transformations of
the Garnier system ${\cal H}_N(\vec{\kappa})$;
then see that it forms an infinite group which contains
a translation ${\mathbb Z}$.

It is known that ${\cal H}_N$ has a symmetry 
which is isomorphic to the symmetric group.
\begin{thm}[{\rm see \cite{IKSY,Ki}}\bf]  \label{thm:sn+3}
The Garnier system
${\cal H}_N(\vec{\kappa})$ 
has birational canonical transformations
\[
\sigma_m:(q,p,s,\vec{\kappa})\mapsto (Q,P,S,\sigma_m(\vec{\kappa})), \quad
1 \leq m \leq N+2,
\]
given in the following table{\rm:}
\[
\begin{array}{|c|c|c|c|c|}
\hline 
\sigma_m  & {\rm action \  on \ } \vec{\kappa}   &  Q_i   & P_i  &S_i 
\\   
 \hline  
\begin{array}{c}
\sigma_{m} \\
(m \leq N)  
\end{array}    
& \theta_m \leftrightarrow \kappa_0  
& 
\begin{array}{l}
Q_i= \displaystyle{\frac{q_i}{R_{im}}} \quad (i \neq m),\\ 
Q_m =  \displaystyle{ \frac{s_m(1-g_s)}{s_m-1}}
\end{array}   
&  
 \begin{array}{l}
P_i= \displaystyle{R_{im} \left(p_i-\frac{s_m}{s_i}p_m \right) },\\ 
P_m = - (s_m-1)p_m
\end{array} 
& 
\begin{array}{l}
S_i= \displaystyle{ \frac{s_m-s_i}{s_m-1}},\\ 
S_m = \displaystyle{ \frac{s_m}{s_m-1}} 
\end{array} 
\\
\hline
\sigma_{N+1}     & \kappa_1 \leftrightarrow \kappa_0  
& \displaystyle Q_i = \frac{q_i}{s_i}    
& P_i =  s_i p_i
& \displaystyle  S_i= \frac{1}{s_i}
\\    \hline
\sigma_{N+2}       & \kappa_1 \leftrightarrow \kappa_\infty   &  Q_i =\displaystyle{ \frac{q_i}{g_1-1}} 
&\begin{array}{l} 
P_i = (g_1-1) \\
\quad \quad
\times
\left(p_i-\alpha-\displaystyle{\sum_j} q_j p_j  \right)
\end{array}
& S_i=  \displaystyle{\frac{s_i}{s_1-1}}
\\
\hline  
\end{array}
\]
where 
$g_1 = \sum_j q_j$,
$g_s = \sum_j  {q_j}/{s_j}$,
and
$\langle \sigma_1,\ldots,\sigma_{N+2} \rangle
\simeq {\frak S}_{N+3}$.
\end{thm}
Theorem~\ref{thm:sn+3} is verified by considering a permutation among $N+3$ singularities of the associated linear differential equation;
see \cite{IKSY,Ki}.
Combine the above ${\frak S}_{N+3}$-symmetry with
the fact that
Hamiltonians $H_i$
(see  (\ref{eq:hamiltonianofgn}))
are invariant under the action
\[
\kappa_\infty \mapsto -\kappa_\infty,
\]
we obtain also the following birational transformations.
\begin{thm}  \label{thm:sym1}
The Garnier system 
${\cal H}_N(\vec{\kappa})$
has the birational canonical transformations
\[  R_{\Delta}:
{\cal H}_N(\vec{\kappa}) \to  {\cal H}_N(R_{\Delta}(\vec{\kappa})).
\]
Here the birational transformations $R_{\Delta}:(q,p) \mapsto (Q,P)$ are described as follows{\rm:} 
\[
\begin{array}{|l|c|l|l|}
\hline 
R_{\Delta} & {\rm action \  on \ } \vec{\kappa}  & \quad Q_i   & \qquad \qquad  \qquad P_i   
\\     \hline  
 R_{\kappa_\infty}     & \kappa_\infty \mapsto -\kappa_\infty  &  Q_i = q_i     & P_i = p_i \\    
 R_{\kappa_1}          & \kappa_1 \mapsto -\kappa_1  &  Q_i = q_i & P_i = p_i - \displaystyle{\frac{\kappa_1}{g_1-1}} \\ 
R_{\kappa_0}          & \kappa_0 \mapsto -\kappa_0  &  Q_i = q_i & P_i = p_i - \displaystyle{\frac{\kappa_0}{s_i(g_s-1)}} \\  
 R_{\theta_j}   & \theta_j \mapsto - \theta_j & Q_i = q_i & P_j = p_j - \frac{\theta_j}{q_j}, \quad  P_i = p_i \quad (i \neq j)  \\ 
\hline  
\end{array}
\]
\end{thm}

We now introduce 
another birational transformation of 
${\cal H}_N(\vec{\kappa})$
which seems to be more nontrivial  than the previous ones.
\begin{thm}     \label{thm:sym2}
The Garnier system 
${\cal H}_N(\vec{\kappa})$
has the birational canonical transformation 
\[ R_{\tau}:
{\cal H}_N(q,p,s, H ;\vec{\kappa})  \to  {\cal H}_N(Q,P,s,\widetilde{H};R_{\tau}(\vec{\kappa})),
\]
where 
$R_{\tau}(\vec{\kappa}) = (-\kappa_0+1,-\kappa_1+1,-\kappa_\infty,-\theta_1,\ldots, -\theta_N  )$
and 
\begin{subequations}
\begin{eqnarray}
Q_i &=& \frac{s_i p_i(q_i p_i - \theta_i)}{\left(\alpha + \sum_jq_j p_j \right)\left(\alpha+\kappa_\infty +  \sum_j q_j p_j \right)}   ,
\label{eq:sym1} \\
Q_i P_i &=& - q_i p_i, 
\label{eq:sym2}  \\
\widetilde{H}_i &=& H_i - \frac{q_i p_i}{s_i}.
\label{eq:sym3}
\end{eqnarray}  
\end{subequations}
\end{thm}

Let $G$ be a group of 
birational canonical transformations of
${\cal H}_N(\vec{\kappa})$ 
defined by
\begin{equation}
{G} = \langle  \sigma_1, \ldots, \sigma_{N+2}, R_{\kappa_0},  R_{\kappa_1},   R_{\kappa_{\infty}}, R_{\theta_1} , \ldots , R_{\theta_N}  ,R_\tau \rangle.
\end{equation}
We see that 
$G$ forms an infinite group which contains ${\mathbb Z}$.
For instance, 
let
\[
l =  R_{\kappa_1}  \circ R_\tau \circ 
R_{\theta_1} \circ  \cdots \circ  R_{\theta_N} \circ R_{\kappa_{\infty}}  \circ R_{\kappa_0} \in {G},
\]
then $l$ acts on the parameter 
as its translation:
\[
l( \vec{\kappa}) = \vec{\kappa} +(1,-1,0,0, \ldots,0),
\]
thus 
$\{ l^n \}  \simeq  {\mathbb Z} 
\subset   {G}$.

\begin{remark} \rm
Group ${G}$ 
might not fill all the birational symmetries of 
${\cal H}_N$.
If $\theta_i=0$ $(i \neq 1)$,
then
${\cal H}_N$ admits a particular solution written in terms of 
solutions of the sixth Painlev\'e equation $P_{\rm VI}$;
see \cite[Theorem~6.1]{T1}.
However
group 
${G}$ 
with the restriction to
$\theta_i=0$ $(i \neq 1)$ 
does not form
the affine Weyl group of type $D_4^{(1)}$,
which is the group of birational symmetries for 
$P_{\rm VI}$ ; 
see \cite{O}.
So the author suspects that
there would exist another hidden symmetry 
of ${\cal H}_N$.
Anyway, 
it is an important problem to 
determine
the group of all birational symmetries 
of the Garnier system 
${\cal H}_N$.
\end{remark}

\noindent
{\it Proof of Theorem \ref{thm:sym2}}. 
First we shall verify that the transformation $R_\tau$ is a canonical transformation of Hamiltonian system  ${\cal H}_N$;
that is,
\begin{equation}  \label{eq:2form}
\sum_i \left( {\rm d}p_i \wedge {\rm d}q_i
  -{\rm d}H_i \wedge {\rm d}s_i \right)=
\sum_i \left( {\rm d}P_i \wedge {\rm d}Q_i -{\rm d} \widetilde{H}_i \wedge {\rm d}s_i \right).
\end{equation}
From (\ref{eq:sym2}),
we have
\begin{equation} \label{eq:dsym2}
P_i {\rm d}Q_i+Q_i{\rm d}P_i
= -p_i{\rm d}q_i-q_i{\rm d}p_i.
\end{equation}
Consider the logarithmic derivative of (\ref{eq:sym1}), 
we have 
\begin{eqnarray} 
\frac{{\rm d}Q_i}{Q_i} &=& \frac{{\rm d}s_i}{s_i} + \frac{{\rm d} p_i}{p_i} + \frac{p_i{\rm d}q_i+q_i{\rm d}p_i }{ q_i p_i - \theta_i} 
\nonumber \\
&& -\left(
\frac{1}{ \alpha + \sum_j q_j p_j}
 +\frac{1}{\alpha+\kappa_\infty + \sum_j q_j p_j } 
\right)    
   \sum_j  {\rm d}(q_jp_j) .
    \label{eq:dsym1}
\end{eqnarray}
By taking the wedge product of (\ref{eq:dsym2})  and (\ref{eq:dsym1}),
we obtain
\begin{eqnarray*}
{\rm d}P_i \wedge {\rm d}Q_i &=& {\rm d}p_i \wedge {\rm d}q_i
-{\rm d} \left(  \frac{q_ip_i}{s_i}  \right) \wedge {\rm d}s_i
\nonumber \\
&& + \left( \frac{1}{ \alpha + \sum_j q_j p_j}
   +\frac{1}{\alpha+\kappa_\infty + \sum_j q_j p_j }\right)
   {\rm d}(q_ip_i) \wedge \sum_{j(\neq i)}  {\rm d}(q_jp_j);
\end{eqnarray*}
hence
\begin{equation}  \label{eq:dpdq}
\sum_i {\rm d}P_i \wedge {\rm d}Q_i 
= \sum_i {\rm d}p_i \wedge {\rm d}q_i
  -\sum_i {\rm d} \left(  \frac{q_ip_i}{s_i}  \right) \wedge {\rm d}s_i.
\end{equation}
On the other hand, it follows from  (\ref{eq:sym3}) that
\begin{equation}  \label{eq:dhds}
{\rm d}\widetilde{H}_i  \wedge {\rm d}s_i
=
{\rm d}H_i  \wedge {\rm d}s_i 
-{\rm d} \left( \frac{q_ip_i}{s_i} \right)   \wedge {\rm d}s_i.
\end{equation}
Combining (\ref{eq:dpdq}) and (\ref{eq:dhds}), we get (\ref{eq:2form}).

Secondly we shall prove that
\begin{equation}  \label{eq:h=h}
\widetilde{H}_i=H_i(Q,P,s,R_\tau(\vec{\kappa})).
\end{equation}
Notice that
$s_j S_{ij} = s_i R_{ji}$.
By using (\ref{eq:sym1}) and (\ref{eq:sym2})  
we have the formulae:
\begin{subequations}  \label{subeq:h=h}
\begin{eqnarray}
&& Q_i \left(-\alpha +\sum_j Q_j P_j \right) \left(-\alpha-\kappa_\infty +  \sum_j Q_j P_j \right)
= s_ip_i(q_ip_i-\theta_i), 
\\
&&   s_iP_i(Q_iP_i+\theta_i) 
= q_i \left(\alpha +\sum_j q_j p_j\right)
        \left(\alpha+\kappa_\infty 
        + \sum_j q_j p_j \right),
\\
&&   \sum_{j(\neq i)} R_{ji}(Q_jP_j+\theta_j)Q_iP_j 
 = \sum_{j(\neq i)} S_{ij}(q_ip_i-\theta_i)q_jp_i ,
\\ 
&& \sum_{j(\neq i)} S_{ij}(Q_iP_i+\theta_i)Q_jP_i
=
 \sum_{j(\neq i)} R_{ji}(q_jp_j-\theta_j)q_ip_j .
\end{eqnarray}
\end{subequations}
Recall the definition of Hamiltonian $H_i$;
see (\ref{eq:hamiltonianofgn}). 
Then we verify (\ref{eq:h=h})  
by (\ref{subeq:h=h}) immediately.
The proof is now complete.
\qed

\section{Toda equation}  \label{sect:toda}

In this section
we show that
a certain sequence of $\tau$-functions 
satisfies the Toda equation.

Since the 1-form  
$\omega =\sum_i H_i {\rm d} s_i$  
is closed,
we can define, up to multiplicative constants, a function $\tau=\tau(s; \vec{\kappa})$ called the 
{\it $\tau$-function} 
by
(see \cite{IKSY,KiO})
\begin{equation}
 {\rm d} \log \tau=\sum_i H_i {\rm d} s_i.
\end{equation}

Let $l$ be a birational canonical transformation of ${\cal H}_N$
defined by
\begin{equation}  \label{eq:defl}
l =  R_{\kappa_1}  \circ R_\tau \circ 
R_{\theta_1} \circ  \cdots \circ  R_{\theta_N} \circ R_{\kappa_{\infty}}  \circ R_{\kappa_0} ,
\end{equation}
then  $l$ acts on the parameter   
$\vec{\kappa} =(\kappa_0 ,\kappa_1, \kappa_{\infty},\theta_1,\ldots ,\theta_N)$
as its translation:
\[
 l( \vec{\kappa}) = \vec{\kappa} +(1,-1,0,0, \ldots,0).
\]
Let 
$(q_i(s),p_i(s),H_i(s))$
be
a solution of the Garnier system 
${\cal H}_N(\vec{\kappa})$
and set
\begin{equation}
\begin{array}{l}
(q_i^+,p_i^+,H_i^+)=(l(q_i),l(p_i),l(H_i)), \\ 
(q_i^-,p_i^-,H_i^-)=(l^{-1}(q_i),l^{-1}(p_i),l^{-1}(H_i)),
\end{array} 
\end{equation}
then we have the

\begin{prop}  \label{prop:H}
The triple of  Hamiltonians
$(H^+_i(s), H_i(s), H^-_i(s))$ 
satisfies 
the differential  equation{\rm:}
\begin{equation} 
        \label{eq:ham}
H^+_i - 2 H_i + H^-_i 
= \frac{\partial}{\partial s_i} \log F(s),
\end{equation}
where
\begin{equation}  \label{eq:f(s)}
F(s) = \left( \sum_j (s_j-1)  \frac{\partial}{\partial s_j} -1\right) 
\sum_ks_k(s_k-1)H_k  
-\kappa_1(\kappa_0-1)+\alpha(\alpha+\kappa_\infty).
\end{equation}
\end{prop}
 
One can prove the proposition by straightforward computations,
via the birational  transformations  given in Sect.~\ref{sect:birat};
see \cite{T6}, for details.

Let 
$\tau^\pm = l^{\pm 1}(\tau)$,
then 
we rewrite
(\ref{eq:ham}) into 
\begin{equation}  \label{eq:toda}
\left(\sum_i (s_i-1)  \frac{\partial}{\partial s_i} -1\right)
\left(\sum_j  s_j(s_j-1) \frac{\partial}{\partial s_j} \right)
\log \tau -\kappa_1(\kappa_0-1)+\alpha(\alpha+\kappa_\infty)
= c \frac{\tau^+  \tau^-}{\tau^2},
\end{equation}
where 
$c$ is a nonzero constant.
Consider the change of variables 
$s_i = {\xi_i}/{(\xi_i-1)}$
and the differential operators: 
\begin{equation}
 A=\sum_i\xi_i \frac{\partial}{\partial \xi_i}, 
\quad
B=\sum_i  \frac{\partial}{\partial \xi_i},
\end{equation}
then we have
\begin{equation}
 \left(\sum_i (s_i-1)  \frac{\partial}{\partial s_i} -1\right)
\left(\sum_j s_j(s_j-1) \frac{\partial}{\partial s_j} \right)  
= (A-B+1)A.
\end{equation}
Note that 
\begin{equation}  \label{eq:AB}
[A,B]=AB-BA=-B.
\end{equation}
Let
\begin{equation}
\psi=\Delta^{\frac{2}{N(N -1)}},
\end{equation} 
where $\Delta$ denotes the difference product of $(\xi_1,\xi_2,\ldots,\xi_N)$, 
{\it i.e.}, 
\[
\Delta = \prod_{i>j} (\xi_i-\xi_j) 
= \left|
\begin{array}{llll}
1 & 1 & \cdots &1 \\
\xi_1 & \xi_2 & \cdots &\xi_N \\
\vdots&\vdots   &  \ddots      &\vdots      \\
\xi_1^{N-1} & \xi_2^{N-1} & \cdots & \xi_N^{N-1}
\end{array}
\right|.
\]
Since 
\[
A \Delta = \frac{N(N-1)}{2} \Delta, \quad
B \Delta = 0,
\]
we have
\begin{equation}    \label{eq:psi}
A \psi = \psi , \quad B \psi = 0.
\end{equation}
Introduce the vector fields
\begin{equation}
X= \psi(A-B), \quad  Y=  \psi A. 
\end{equation}
One can easily verify that $[X,Y]=0$, 
\begin{equation}
XY=\psi^2(A-B+1)A,  \label{eq:XY}
\end{equation}
and
\begin{equation}
XY \log \psi=\psi^2.   \label{eq:XYlog}
\end{equation}
by using (\ref{eq:AB}) and (\ref{eq:psi}).

Let us consider the sequence of $\tau$-functions
$\{\tau_n | n\in{\mathbb Z} \}$ defined by
\begin{equation} \label{eq:tauseq}
\tau_n = \psi^{a_n} l^n(\tau),
\end{equation}
with
\begin{equation}
a_n =  -(\kappa_1-n)(\kappa_0+n-1)+\alpha(\alpha+\kappa_\infty).
\end{equation} 
Substitute (\ref{eq:tauseq})
into (\ref{eq:toda}),  
by virtue of (\ref{eq:XY})  and (\ref{eq:XYlog}),
we now arrive at the

\begin{thm}   \label{thm:toda}
The
sequence 
$\{\tau_n |n\in{\mathbb Z}\}$
satisfies the  Toda equation{\rm:}
\begin{equation}   \label{eq:thmtoda}
X Y \log \tau_n = c_n \frac{ \tau_{n-1} \tau_{n+1}}{\tau_n^2}, 
\end{equation}
where $X$, $Y$ 
being vector fields such that $[X,Y]=0$
and $c_n$  a nonzero constant. 
\end{thm}

\begin{remark} \rm
A sequence of $\tau$-functions corresponding to
other translations also 
satisfies the Toda equation.
For instance, 
let us consider  the birational transformation $\widetilde{l}$ 
defined by
\begin{equation}  \label{eq:defltilde}
\widetilde{l} = R_{\kappa_1} \circ l \circ R_{\kappa_1},
\end{equation}
which acts on the parameter $\vec{\kappa}$ as 
its translation:
\[
\widetilde{l}( \vec{\kappa}) = \vec{\kappa} +(1,1,0,0, \ldots,0).
\]
It is easy to see that
\begin{equation}
R_{\kappa_1}(\tau) = \tau  \prod_i (s_i-1)^{-\kappa_1 \theta_i} .
\end{equation}
Combine this with 
(\ref{eq:toda}), 
we obtain 
\begin{equation}  \label{eq:toda2}
\left(\sum_i  \frac{\partial}{\partial s_i} -1\right)
\left(\sum_j s_j(s_j-1) \frac{\partial}{\partial s_j} \right)
\log \tau +\alpha(\alpha+\kappa_\infty)
= c \frac{ \widetilde{l}^{-1}(\tau) \, \widetilde{l}(\tau) }{\tau^2}.
\end{equation}
Also
(\ref{eq:toda2}) is equivalent to the Toda equation
via a similar change of variables as above.
\end{remark}

\section{Algebraic solutions in terms of universal characters}  \label{sect:alg}

In this section
we construct
a class of algebraic solutions of the Garnier system ${\cal H}_N$ and
then express it in terms of the universal characters.

\subsection{Algebraic solutions}
Consider the birational canonical transformation
\begin{equation}
w_0=R_{\tau} \circ R_{\theta_1} \circ \cdots \circ  R_{\theta_N} \circ  R_{\kappa_\infty},
\end{equation}
given as follows:
\[ w_0:
{\cal H}_N(q,p;\vec{\kappa})  \to  {\cal H}_N(Q,P;w_0(\vec{\kappa})),
\]
where $w_0(\vec{\kappa}) = (-\kappa_0+1,-\kappa_1+1,\kappa_\infty,\theta_1,\ldots,\theta_N)$ and 
\begin{subequations}
\begin{eqnarray}
Q_i &=& \frac{s_i p_i(q_i p_i - \theta_i)} {\left(\alpha + \sum_j q_j p_j
\right)
\left(\alpha+\kappa_\infty +  \sum_j q_j p_j\right) } ,\\
Q_i P_i &=& - q_i p_i + \theta_i.
\end{eqnarray}
\end{subequations}
If $\kappa_0 = \kappa_1 = 1/2$, 
the fixed point with respect to the action of $w_0$
is 
\begin{equation}  \label{eq:seed}
(q_i,p_i)=\left(\frac{\theta_i \sqrt{s_i}}{\kappa_\infty},\frac{\kappa_\infty}{2\sqrt{s_i}} \right), \quad
i=1,\ldots ,N.
\end{equation}
This is an algebraic solution of 
${\cal H}_N$.
Applying the birational symmetries ${G}$ (see Sect.~\ref{sect:birat})
to 
(\ref{eq:seed}),
we obtain a class of algebraic solutions.
\begin{thm}  \label{thm:algsol}
If two components of the parameter 
$\vec{\kappa}=
(\kappa_0,\kappa_1,\kappa_\infty,\theta_1,\ldots, \theta_N)$
are half integers
then ${\cal H}_N$ admits an algebraic solution.
\end{thm}

\subsection{Special polynomials}
Substituting the algebraic solution, (\ref{eq:seed}), 
into Hamiltonians (see (\ref{eq:hamiltonianofgn})),
we have 
\begin{equation}
s_i(s_i-1)H_i = -\frac{1}{2} \kappa_\infty \theta_i \sqrt{s_i}+ \frac{1}{4}\theta_i(s_i-1) + \frac{1}{2}\sum_j \theta_i \theta_j
 \frac{\sqrt{s_is_j}+1 }{ \sqrt{{s_j}/{s_i}}+1 };
\end{equation}
and then the corresponding 
$\tau$-function 
is given as follows:
\begin{equation} \label{eq:tau00}
\tau_{0,0}
=
\prod_i s_i^{-\theta_i(\theta_i-1)/4}
(\sqrt{s_i}+1)^{ \theta_i( \sum_k \theta_k+\kappa_\infty)/2} 
(\sqrt{s_i}-1)^{\theta_i( \sum_k \theta_k-\kappa_\infty)/2}
\prod_{i,j}(\sqrt{s_i}+\sqrt{s_j})^{-\theta_i \theta_j/2 }.
\end{equation}

Let us consider the birational transformations $l$ and $\widetilde{l}$, 
defined  respectively by (\ref{eq:defl}) and (\ref{eq:defltilde}),
which 
act on the parameter $\vec{\kappa}$
as its translations: 
\begin{equation}
\begin{array}{l}
l( \vec{\kappa}) = \vec{\kappa} +(1,-1,0,0, \ldots,0),
\\
\widetilde{l}( \vec{\kappa}) = \vec{\kappa} +(1,1,0,0, \ldots,0).
\end{array}
\end{equation}
Introduce a family of 
$\tau$-functions 
$\tau_{m,n}$ $(m,n \in {\mathbb Z})$ 
defined by
\begin{equation}
{\widetilde l}^m l^n (\tau_{0,0}) = \tau_{m,n}.
\end{equation}
Let
\begin{equation}
s_i=t_i^2,
\end{equation}
then (\ref{eq:tau00}) is rewritten as 
\begin{subequations}   \label{subeq:init}
\begin{equation}
\tau_{0,0} =
\prod_i t_i^{-\theta_i(\theta_i-1)/2}
(t_i+1)^{\theta_i( \sum_k \theta_k+\kappa_\infty)/2} 
(t_i-1)^{\theta_i( \sum_k \theta_k-\kappa_\infty)/2}
\prod_{i,j}(t_i+t_j)^{-{\theta_i \theta_j}/{2} }.
\end{equation}
Applying the action of $\widetilde{l}$ and $l$, 
we see that 
\begin{eqnarray} 
   \tau_{0,1}&=& \prod_i t_i^{-\theta_i} \tau_{0,0}, \\
   \tau_{1,0}&=& \left(\prod_i t_i^{-\theta_i} (t_i+1)^{\theta_i} (t_i-1)^{\theta_i} \right)
\left(  \sum_j \theta_j t_j - \kappa_\infty \right)
\tau_{0,0},
\\
   \tau_{1,1} &=& \left( \prod_i t_i^{-2 \theta_i} (t_i+1)^{\theta_i} (t_i-1)^{\theta_i} \right)
\left(\kappa_\infty - \sum_j \theta_j t_j^{-1}  \right)
\tau_{0,0}.
\end{eqnarray}
\end{subequations}
The $\tau$-functions, $\tau_{m,n}$ $(m,n \in {\mathbb Z})$,
are determined successively by the use of  
the Toda equations
(\ref{eq:toda}) and (\ref{eq:toda2}),
from the above initial values (\ref{subeq:init}).

Now let us define the functions, 
$T_{m,n}=T_{m,n}(t)$ $(m,n \in {\mathbb Z})$,
by 
\begin{eqnarray}
T_{m,n}(t) &=&  \tau_{m,n}
\prod_i 
\biggl\{ t_i^{(\theta_i+m+n)(\theta_i+m+n-1)/2}
(t_i+1)^{-\theta_i( \sum_k \theta_k+\kappa_\infty +2 m)/2} 
\nonumber\\
&& \qquad
\times
(t_i-1)^{-\theta_i( \sum_k \theta_k-\kappa_\infty +2 m)/2}
\biggr\}
\prod_{i,j}(t_i+t_j)^{{\theta_i \theta_j}/{2} }.    
\label{eq:defTmn}
\end{eqnarray}
Substituting (\ref{eq:defTmn}) 
into 
(\ref{eq:toda}) and (\ref{eq:toda2}) 
with $c=1/4$,
we thus obtain the 
recurrence relations 
for 
$T_{m,n}$.

\begin{prop}
The function $T_{m,n}=T_{m,n}(t)$ 
$(m,n \in {\mathbb Z})$
satisfies the following recurrence relations{\rm:}
\begin{subequations}  \label{subeq:Tmn}
\begin{eqnarray}
T_{m+1,n} &=& \prod_i t_i \Biggl\{ 
 \left(\sum_i \frac{t_i^2-1}{t_i}  \frac{\partial}{\partial t_i} -2 \right)
\sum_i t_i(t_i^2-1)  \frac{\partial}{\partial t_i}
       \log T_{m,n} \nonumber \\  
&& +\kappa_\infty \sum_i \theta_i \frac{t_i^2+1}{t_i}
   -\frac{1}{2} \sum_{i,j} \theta_i \theta_j \frac{t_i^2+t_j^2}{t_i t_j} 
-\kappa_\infty^2+(2m)^2 
\Biggr\}   \frac{{T_{m,n}}^2}{T_{m-1,n}}, 
\label{eq:Tmn1}
\\ 
T_{m,n+1} &=&  \prod_i t_i \Biggl\{ 
 \left(\sum_i \frac{t_i^2-1}{t_i}  \frac{\partial}{\partial t_i} -2 \right)
\sum_i t_i(t_i^2-1)  \frac{\partial}{\partial t_i}
       \log T_{m,n} \nonumber \\  
&& +\kappa_\infty \sum_i \theta_i \frac{t_i^2+1}{t_i}
   -\frac{1}{2} \sum_{i,j} \theta_i \theta_j \frac{t_i^2+t_j^2}{t_i t_j} 
-\kappa_\infty^2+(2n-1)^2 
\Biggr\} 
 \frac{{T_{m,n}}^2}{T_{m,n-1}}.
\label{eq:Tmn2}
\end{eqnarray}
\end{subequations}
Here the initial values are given as follows{\rm:}
\begin{equation} \label{eq:T00}
T_{0,0}=T_{0,1}=1, \quad  T_{1,0}=\sum_i \theta_i t_i - \kappa_\infty, 
\quad  T_{1,1}= \prod_i t_i \left(   \kappa_\infty - \sum_j \theta_j t_j^{-1} \right).
\end{equation}
\end{prop}

We call $T_{m,n}(t)$ 
{\it special polynomials}
associated with algebraic solutions of ${\cal H}_N$.
By the above recurrence relations (\ref{subeq:Tmn}),
we can only state that 
$T_{m,n}(t)$
are rational functions in 
$t=(t_1,\ldots,t_N)$.
We will show that 
$T_{m,n}(t)$ are indeed  polynomials;
see Theorem~\ref{thm:Tmnuc} and  Corollary~\ref{cor:Tmnuc} below.
Note that
\begin{equation}   \label{eq:symT}
T_{-m,n}(t)=T_{m,1-n}(t)=
(-1)^{m(2n-1)} \prod_i t_i^{m^2+n(n-1)}
T_{m,n}(t^{-1}),
\end{equation}
which is verified easily
by the recurrence relations and  initial values.
Algebraic solutions of ${\cal H}_N$ are explicitly 
written in terms of the special polynomials $T_{m,n}(t)$.

\begin{thm}  \label{thm:algsolinT}
If 
$\kappa_0 = 1/2+m+n$, 
$\kappa_1 =  1/2+m-n$
$(m,n \in {\mathbb Z})$,
then 
${\cal H}_N$ 
admits an algebraic solution given as follows{\rm:}
\begin{subequations}  \label{subeq:alg}
\begin{eqnarray}
q_i &=&  \frac{ \displaystyle{t_i \frac{\partial}{\partial t_i} \log \frac{T_{m+1,n}}{ T_{m,n+1}} } }
{\displaystyle \sum_j t_j \frac{\partial}{\partial t_j} \log \frac{T_{m+1,n}}{ T_{m,n+1}} 
-2m+2n-1},
\label{eq:alg1}
\\ \nonumber \\
2 q_i p_i &=& \theta_i+m+n + t_i \frac{\partial}{\partial t_i} \log \frac{T_{m,n}}{ T_{m,n+1}} .
\label{eq:alg2}
\end{eqnarray}
\end{subequations}
\end{thm}

\pf
By using the birational canonical transformations
$l$ and $\widetilde{l}$,
we have
\begin{eqnarray}
l(H_i) &=&H_i -\frac{ q_i p_i}{s_i}, \\
\widetilde{l}(H_i) 
&=&
H_i - \frac{1}{s_i} \left(q_i p_i- \frac{\kappa_1 q_i}{g_1-1}\right)
+\frac{\theta_i}{s_i-1},
\end{eqnarray}
where $g_1 = \sum_j q_j$.
We then obtain the relation between
$\tau$-functions and canonical variables:
\begin{subequations}   \label{subeq:taucan}
\begin{eqnarray}
q_i &=&  \frac{ \displaystyle{s_i \frac{\partial}{\partial s_i} \log \frac{ \widetilde{l}(\tau)}{l(\tau) } }  
- \frac{\theta_i s_i}{s_i-1} }
{ \displaystyle \sum_j \left( s_j \frac{\partial}{\partial s_j} 
\log \frac{\widetilde{l}(\tau) }{l(\tau) } -   
\frac{\theta_j s_j}{s_j-1}    \right) -
\kappa_1    },
\label{eq:taucan1}
\\ 
\nonumber \\
q_i \, p_i &=& s_i \frac{\partial}{\partial s_i} \log 
\frac{\tau}{ l(\tau)} .
\label{eq:taucan2}
\end{eqnarray}
\end{subequations}
Here recall the definition of $\tau$-function,
$\partial/\partial s_i  \log \tau =H_i$.
Substitute (\ref{eq:defTmn})
into (\ref{subeq:taucan})
with $s_i=t_i^2$, 
we get
(\ref{subeq:alg}).
\qed

\subsection{Universal characters}
To investigate the special polynomial $T_{m,n}$ in detail,
we have to recall the definition of the universal characters;
see \cite{Ko,T3}.
For each pair of partitions
$[\lambda,\mu] = 
[ (\lambda_1,\lambda_2, \ldots,\lambda_l), (\mu_1,\mu_2, \ldots,\mu_{l'})]$,
the {\it universal character} 
$S_{[\lambda,\mu ]}(x,y)$
is a polynomial in 
$(x,y)=(x_1,x_2,\ldots,y_1,y_2,\ldots)$
defined as follows:
\begin{equation}
S_{[\lambda,\mu ]}(x,y)
= \det 
\left(
  \begin{array}{ll}
 q_{\mu_{l'-i+1}  +i - j }(y),  &  1 \leq i \leq l'  \\
 p_{\lambda_{i-l'}-i+j}(x),     &  l'+1 \leq i \leq l+l'   \\
  \end{array}
\right)_{1 \leq i,j \leq l+l'} .
\end{equation}
Here    
$p_n(x)$
is determined by the generating function:
\begin{equation} \label{eq:elems}
\sum_{n=0}^\infty p_n(x)z^n = e^{\xi(x,z)}, \quad 
\xi(x,z)=\sum_{n=1}^\infty x_n z^n, 
\end{equation}
and set $p_{-n}(x)=0$ for $n>0$;
$q_n(y)$ is the same as $p_n(x)$ 
except
replacing $x$ with $y$.
Note that $p_n(x)$ is explicitly written as follows:
\begin{equation}  
p_n(x)=
\sum_{k_1+2k_2+ \cdots +n k_n=n}
\frac{x_1^{k_1} x_2^{k_2} \cdots x_n^{k_n}}{k_1!k_2! \cdots k_n!}.
\end{equation}
If we count the degree of each variable $x_n$ and $y_n$ $(n=1,2,\ldots)$ 
as
\[ \deg x_n =n \quad
{\rm and}  \quad
 \deg y_n=-n,
\]
then
the universal character $S_{[\lambda,\mu ]}(x,y)$ 
is a weighted homogeneous polynomial of degree $|\lambda| -|\mu|$,
where we let $|\lambda|=\lambda_1+\cdots +\lambda_l$.
Note that 
the Schur polynomial $S_\lambda(x)$ 
(see {\it e.g.}  \cite{Mac})
is regarded as
a special case of the universal character: 
\[
S_\lambda (x) 
= \det \left( p_{\lambda_i-i+j}(x) \right)
= S_{[\lambda, \emptyset ]}(x,y).
\]

\begin{example}   \rm
When $\lambda=(2,1)$, $\mu=(1)$,
the universal character is given as follows: 
\[
S_{[(2,1),(1)]}(x,y) = 
\left|
\begin{array}{ccc}
q_1 & q_0 & q_{-1} \\
p_1 & p_2& p_3 \\
p_{-1}&p_0&p_1
\end{array}
\right|
= y_1 \left(\frac{x_1^3}{3} -x_3  \right)-x_1^2,
\]
which is a weighted homogeneous polynomial of degree $|\lambda|-|\mu|=2$.
\end{example}

The special polynomial $T_{m,n}(t)$ can be written in terms of the universal character.

\begin{thm}  \label{thm:Tmnuc}
The special polynomial $T_{m,n}(t)$
$(m,n \in {\mathbb Z})$ 
is expressed  as follows{\rm:}
\begin{equation}
T_{m,n}(t) = N_{m,n}
S_{[\lambda,\mu]}(x,y).
\end{equation}
Here
$\lambda = (u,u-1, \ldots,2,1)$, 
$\mu=(v,v-1,\ldots,2,1)$
with
$u=|n-m-{1}/{2}|-{1}/{2}$,  
$v=|n+m-{1}/{2}|-{1}/{2}${\rm;}
and 
\begin{equation}  \label{eq:spec}
x_n=\frac{-\kappa_\infty+\sum_{i} \theta_i t_i^n}{n}, 
\quad 
y_n=\frac{-\kappa_\infty+\sum_{i}\theta_i t_i^{-n}}{n}.
\end{equation}
The normalization factor $N_{m,n}$
is given by
\begin{equation}
N_{m,n}=(-1)^{ {v(v+1)}/{2}} 
             \prod_{i=1}^N t_i^{ {v(v+1)}/{2}} 
             \prod_{j=1}^u(2j-1)!!
             \prod_{k=1}^v(2k-1)!!.  
\end{equation}
\end{thm}

Consequently we have the
\begin{cor}  \label{cor:Tmnuc}
The special polynomial $T_{m,n}(t)$ is indeed a polynomial 
of degree 
$m^2+n(n-1)${\rm;}
furthermore 
$T_{m,n}(t) \in {\mathbb Z}[\kappa_\infty,\theta_1,\ldots,\theta_N][t]$.
\end{cor}

The proof of Theorem~\ref{thm:Tmnuc} is given in Sect.~\ref{sect:proof}.

We show in Figure 1 below how 
the special polynomials 
$T_{m,n}(t)$
are arranged on  
$(m,n)$-lattice.
We also give some examples of 
$T_{m,n}(t)$
of small degrees in the case $N=1$.

\begin{center}
\begin{picture}(480,400)
\put(215,100){(0,0)}
\put(275,100){(1,0)}
\put(215,160){(0,1)}
\put(275,160){(1,1)}

\put(90,20){\vector(0,1){20}}
\put(150,20){\vector(0,1){20}}
\put(210,20){\vector(0,1){20}}
\put(270,20){\vector(0,1){20}}
\put(330,20){\vector(0,1){20}}
\put(390,20){\vector(0,1){20}}

\put(50,60){\vector(1,0){20}}
\put(90,60)
{\makebox(0,0){$\emptyset  \, , \, {\scriptstyle \youngc{123}} $}}
\put(110,60){\vector(1,0){20}}
\put(150,60)
{\makebox(0,0){$\emptyset  \, , \, {\scriptstyle \youngc{12}} $}}
\put(170,60){\vector(1,0){20}}
\put(210,60)
{\makebox(0,0){${\scriptstyle \youngc{1}  \, , \, \youngc{1}} $}}
\put(230,60){\vector(1,0){20}}
\put(270,60)
{\makebox(0,0){${\scriptstyle \youngc{12}} \, , \, \emptyset $}}
\put(290,60){\vector(1,0){20}}
\put(330,60)
{\makebox(0,0){${\scriptstyle \youngc{123}} \, , \, \emptyset $}}
\put(350,60){\vector(1,0){20}}
\put(390,60)
{\makebox(0,0){${\scriptstyle \youngc{1234} \, , \, \youngc{1}} $}}
\put(410,60){\vector(1,0){20}}

\put(90,80){\vector(0,1){20}}
\put(150,80){\vector(0,1){20}}
\put(210,80){\vector(0,1){20}}
\put(270,80){\vector(0,1){20}}
\put(330,80){\vector(0,1){20}}
\put(390,80){\vector(0,1){20}}

\put(50,120){\vector(1,0){20}}
\put(90,120)
{\makebox(0,0){${\scriptstyle \youngc{1}  \, , \, \youngc{12}} $}}
\put(110,120){\vector(1,0){20}}
\put(150,120)
{\makebox(0,0){$\emptyset  \, , \, {\scriptstyle \youngc{1}} $}}
\put(170,120){\vector(1,0){20}}
\put(210,120)
{\makebox(0,0){$\emptyset  \, , \, \emptyset $}}
\put(230,120){\vector(1,0){20}}
\put(270,120)
{\makebox(0,0){${\scriptstyle \youngc{1}} \, , \, \emptyset $}}
\put(290,120){\vector(1,0){20}}
\put(330,120)
{\makebox(0,0){${\scriptstyle \youngc{12} \, , \, \youngc{1} } $}}
\put(350,120){\vector(1,0){20}}
\put(390,120)
{\makebox(0,0){${\scriptstyle \youngc{123} \, , \, \youngc{12} } $}}
\put(410,120){\vector(1,0){20}}

\put(90,140){\vector(0,1){20}}
\put(150,140){\vector(0,1){20}}
\put(210,140){\vector(0,1){20}}
\put(270,140){\vector(0,1){20}}
\put(330,140){\vector(0,1){20}}
\put(390,140){\vector(0,1){20}}

\put(50,180){\vector(1,0){20}}
\put(90,180)
{\makebox(0,0){${\scriptstyle \youngc{12} \, , \, \youngc{1}} $}}
\put(110,180){\vector(1,0){20}}
\put(150,180)
{\makebox(0,0){${\scriptstyle \youngc{1}} \, , \, \emptyset $}}
\put(170,180){\vector(1,0){20}}
\put(210,180)
{\makebox(0,0){$\emptyset \, , \, \emptyset $}}
\put(230,180){\vector(1,0){20}}
\put(270,180)
{\makebox(0,0){$\emptyset \, , \,{\scriptstyle \youngc{1}} $}}
\put(290,180){\vector(1,0){20}}
\put(330,180)
{\makebox(0,0){${\scriptstyle \youngc{1} \, , \, \youngc{12} } $}}
\put(350,180){\vector(1,0){20}}
\put(390,180)
{\makebox(0,0){${\scriptstyle \youngc{12} \, , \, \youngc{123} } $}}
\put(410,180){\vector(1,0){20}}

\put(90,200){\vector(0,1){20}}
\put(150,200){\vector(0,1){20}}
\put(210,200){\vector(0,1){20}}
\put(270,200){\vector(0,1){20}}
\put(330,200){\vector(0,1){20}}
\put(390,200){\vector(0,1){20}}

\put(50,240){\vector(1,0){20}}
\put(90,240)
{\makebox(0,0){${\scriptstyle \youngc{123}} \, , \emptyset $}}
\put(110,240){\vector(1,0){20}}
\put(150,240)
{\makebox(0,0){${\scriptstyle \youngc{12}} \, , \, \emptyset $}}
\put(170,240){\vector(1,0){20}}
\put(210,240)
{\makebox(0,0){${\scriptstyle \youngc{1} \, , \, \youngc{1} } $}}
\put(230,240){\vector(1,0){20}}
\put(270,240)
{\makebox(0,0){$\emptyset \, , \,{\scriptstyle \youngc{12}} $}}
\put(290,240){\vector(1,0){20}}
\put(330,240)
{\makebox(0,0){$\emptyset \, , \,{\scriptstyle \youngc{123}} $}}
\put(350,240){\vector(1,0){20}}
\put(390,240)
{\makebox(0,0){${\scriptstyle \youngc{1} \, , \, \youngc{1234}} $}}
\put(410,240){\vector(1,0){20}}

\put(90,260){\vector(0,1){20}}
\put(150,260){\vector(0,1){20}}
\put(210,260){\vector(0,1){20}}
\put(270,260){\vector(0,1){20}}
\put(330,260){\vector(0,1){20}}
\put(390,260){\vector(0,1){20}}

\put(50,300){\vector(1,0){20}}
\put(90,300)
{\makebox(0,0){${\scriptstyle \youngc{1234}} \, , \emptyset $}}
\put(110,300){\vector(1,0){20}}
\put(150,300)
{\makebox(0,0){${\scriptstyle \youngc{123} \, , \, \youngc{1} }$}}
\put(170,300){\vector(1,0){20}}
\put(210,300)
{\makebox(0,0){${\scriptstyle \youngc{12} \, , \, \youngc{12} } $}}
\put(230,300){\vector(1,0){20}}
\put(270,300)
{\makebox(0,0){${\scriptstyle \youngc{1}  \, , \, \youngc{123}} $}}
\put(290,300){\vector(1,0){20}}
\put(330,300)
{\makebox(0,0){$\emptyset \, , \,{\scriptstyle \youngc{1234}} $}}
\put(350,300){\vector(1,0){20}}
\put(390,300)
{\makebox(0,0){$\emptyset \, , \,{\scriptstyle \youngc{12345}} $}}
\put(410,300){\vector(1,0){20}}

\put(90,320){\vector(0,1){20}}
\put(150,320){\vector(0,1){20}}
\put(210,320){\vector(0,1){20}}
\put(270,320){\vector(0,1){20}}
\put(330,320){\vector(0,1){20}}
\put(390,320){\vector(0,1){20}}

\put(50,360){\vector(1,0){20}}
\put(90,360)
{\makebox(0,0){${\scriptstyle \youngc{12345} \, , \, \youngc{1} } $}}
\put(110,360){\vector(1,0){20}}
\put(150,360)
{\makebox(0,0){${\scriptstyle \youngc{1234} \, , \, \youngc{12} }$}}
\put(170,360){\vector(1,0){20}}
\put(210,360)
{\makebox(0,0){${\scriptstyle \youngc{123} \, , \, \youngc{123} } $}}
\put(230,360){\vector(1,0){20}}
\put(270,360)
{\makebox(0,0){${\scriptstyle \youngc{12}  \, , \, \youngc{1234}} $}}
\put(290,360){\vector(1,0){20}}
\put(330,360)
{\makebox(0,0){${\scriptstyle \youngc{1} \, , \, \youngc{12345}} $}}
\put(350,360){\vector(1,0){20}}
\put(390,360)
{\makebox(0,0){$\emptyset \, , \,{\scriptstyle  \youngc{123456}} $}}
\put(410,360){\vector(1,0){20}}

\put(90,380){\vector(0,1){20}}
\put(150,380){\vector(0,1){20}}
\put(210,380){\vector(0,1){20}}
\put(270,380){\vector(0,1){20}}
\put(330,380){\vector(0,1){20}}
\put(390,380){\vector(0,1){20}}

\end{picture}
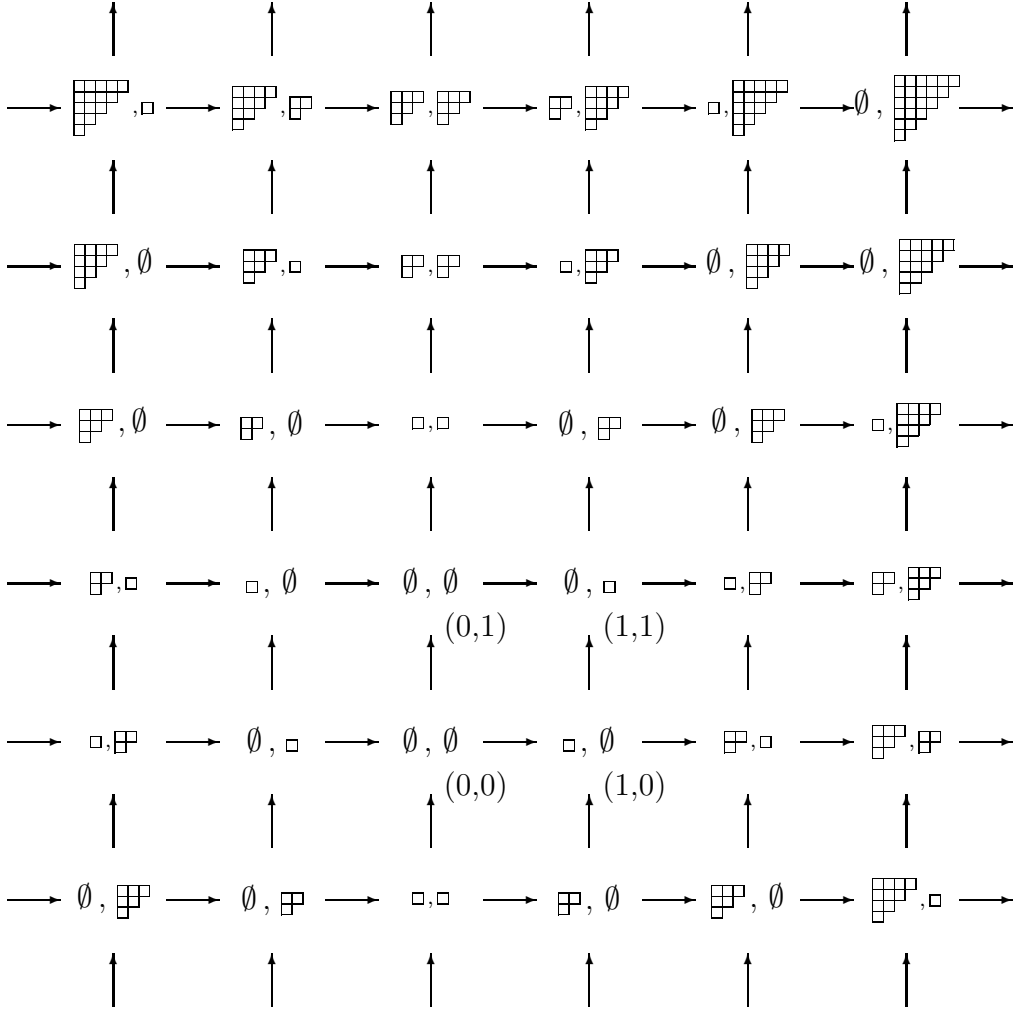
{\bf Figure 1} 
Special polynomials $T_{m,n}(t)$.
\end{center}

\noindent
The special polynomials $T_{m,n}(t)$ for $N=1$ are as follows:
\begin{eqnarray*}
&&T_{0,0}=T_{0,1}=1, \quad 
T_{1,0}=T_{-1,1}=-\kappa_\infty+\theta t, \quad
T_{1,1}=T_{-1,0}= - \theta + \kappa_\infty t, 
\\
&&
T_{0,2}=T_{0,-1}
= \kappa_\infty \theta + t -\kappa_\infty^2 t-\theta^2 t+\kappa_\infty \theta t^2,  \\
&&
T_{1,-1}=T_{-1,2}= 
\kappa_\infty - \kappa_\infty^3 + 3 \kappa_\infty^2 \theta t
-  3 \kappa_\infty \theta^2 t^2 - \theta t^3 + \theta^3 t^3, 
\\
&&
T_{1,2}=T_{-1,-1}=
\theta-\theta^3+3 \kappa_\infty \theta^2 t 
- 3 \kappa_\infty^2 \theta t^2 - \kappa_\infty t^3 +\kappa_\infty^3 t^3, 
\\
&&
T_{2,0} = T_{-2,1} =
-\kappa_\infty \theta + \kappa_\infty^3 \theta + 4 \kappa_\infty^2 t - \kappa_\infty^4 t - 
    3 \kappa_\infty^2 \theta^2 t - 6 \kappa_\infty \theta t^2 + 
    3 \kappa_\infty^3 \theta t^2 + 3 \kappa_\infty \theta^3 t^2 \\
&& \qquad \qquad   \qquad   \quad
+ 4 \theta^2 t^3 - 
    3 \kappa_\infty^2 \theta^2 t^3 - \theta^4 t^3 - \kappa_\infty \theta t^4 + 
    \kappa_\infty \theta^3 t^4.
\end{eqnarray*}

\begin{remark}  \rm
Under the specialization (\ref{eq:spec}),
we let $p_n(x)=P_n(t)$.
Then the generating function (\ref{eq:elems}) 
is rewritten as follows:
\begin{equation}
\sum_{n=0}^\infty P_n(t) z^n=
(1-z)^{\kappa_\infty} \prod_i(1-t_i z)^{-\theta_i}.
\end{equation}
Hence
$P_n(t)$ has the following expression:
\begin{equation}
P_n(t)=
\frac{(-\kappa_\infty)_n}{(1)_n}
  F_D(-n, \theta_1, \ldots, \theta_N,\kappa_\infty -n+1;t),
\end{equation}
where $F_D$ denotes the Lauricella hypergeometric series
and 
$(a)_n=a (a+1)(a+2)\cdots(a+n-1)$;
see {\it e.g.} \cite{IKSY,OKi,T2}.
\end{remark}

\begin{remark}  \rm
If $N=1$, $T_{m,n}(t)$ is equivalent to the {\it Umemura polynomial} of $P_{\rm VI}$,
for which Masuda considered its explicit formula in terms of universal characters;
see \cite{M,NOOU}.
We refer also to the results \cite{MOK} and \cite{T4},
where a class of rational  solutions of 
$P_{\rm V}$ and that of the
(higher order) Painlev\'e equation of type $A^{(1)}_{2g+1}$ $(g\geq 1)$
are obtained in terms of universal characters.
\end{remark}

\begin{remark}  \rm
Several other classes of solutions of the Garnier system have been studied. 
In \cite{T2}, a family of rational solutions  was obtained 
by the use of
Schur polynomials.
In \cite{KK}, 
solutions in terms of hyperelliptic theta functions were considered 
from the viewpoint of algebraic geometry.
\end{remark}

\section{Proof of Theorem~\ref{thm:Tmnuc}}
\label{sect:proof}

\subsection{A generalization of Jacobi's identity}
First we prepare an identity 
for determinants,
which is regarded as a generalization of Jacobi's identity.
Let $A=(a_{ij})_{i,j}$ be an $n \times n$ matrix and 
$\xi^I_J= \xi^I_J(A)$ its minor determinant 
with respect to
rows $I=\{i_1,\ldots,i_r\}$ and 
columns $J=\{j_1,\ldots,j_r\}$.
For two disjoint sets
$I,J \subset \{1,\ldots,n\}$,
we define 
$\epsilon(I;J)$ by
\begin{equation}
\epsilon(I;J)=(-1)^{l(I;J)}, \quad
l(I;J)= \#  \left\{(i,j) \in I \times J  \,  |  \, i>j   \right\}.
\end{equation}

\begin{thm}  \label{thm:jacobi}
Let $I=\{1,2,\ldots,n \}$ and 
$A=(a_{ij})_{i,j\in I}$.
The following  quadratic
relation
among minor determinants of $A$ holds{\rm:}
\begin{equation}  \label{eq:jacobi}
\xi^I_I \xi^{I- J_1-J_2}_{I-J_1-J_2}=
\sum_{\begin{subarray}{c} 
K_1,K_2 \subset I;\\
K_1 \cap (I-J_1-J_2) = \emptyset;\\
K_2\cap (I-J_1-J_2) = \emptyset
\end{subarray}}  
\epsilon(K_1;K_2) 
\xi^{I-K_1}_{I-J_1} \xi^{I-K_2}_{I-J_2},
\end{equation}
where $|J_1|=|K_1|=r_1$ and  
$|J_2|=|K_2|=r_2$.
\end{thm}

Let $r_1=r_2=1$,
$J_1 =\{1 \}$ and 
$J_2 =\{n \}$,
then 
(\ref{eq:jacobi}) 
recovers  Jacobi's identity 
(see \cite{J}):  
\begin{equation}
\xi^{1 \cdots n}_{1 \cdots n} 
\xi^{2 \cdots n-1}_{2 \cdots n-1} 
= \xi^{2 \cdots n}_{2 \cdots n} 
\xi^{1 \cdots n-1}_{1 \cdots n-1} -
\xi^{1 \cdots n-1}_{2 \cdots n} 
\xi^{2 \cdots n}_{1 \cdots n-1},
\end{equation}
in fact.

\noindent
{\it Proof of Theorem \ref{thm:jacobi}.}
Without loss of generality,
we can set 
$J_1= \{ 1,2,\ldots,r_1 \}$
and
$J_2= \{ n-r_2+1,\ldots,n-1,n \}$.
Let
$\widetilde{I} = \{ 1,2, \ldots, 2n-r_1-r_2 \}$.
Consider a
$(2n-r_1-r_2) \times (2n-r_1-r_2)$ matrix 
$B=(b_{ij})_{i,j \in \widetilde{I}}$
given as follows: 
\begin{equation}
\begin{array}{llll}
(\rm i)&b_{ij}=a_{ij}& {\rm for} &
i,j \in I; \\
(\rm ii)& b_{ij}=a_{i,j-n+r_1}& {\rm for} &
i \in I, \
j \in \widetilde{I} \setminus I; \\
(\rm iii)& b_{ij}=a_{i-n+r_1,j} &{\rm for} & 
i \in \widetilde{I} \setminus I,  \
j \in J_1; \\
(\rm iv)& b_{ij}=0 &{\rm for} &
i \in \widetilde{I} \setminus I,  \
j \in I \setminus J_1;  \\
(\rm v) &b_{ij}=a_{i-n+r_1,j-n+r_1}  &{\rm for} &
i \in \widetilde{I} \setminus I,  \
j \in \widetilde{I} \setminus I,
\end{array}
\end{equation}
{\it i.e.},
write $A$ as 
\[
A =\left[ 
\begin{array}{c|c|c}
A_{11} & A_{12} & A_{13} \\
\hline
A_{21} & {\boldsymbol A}_{22} & A_{23} \\
\hline
A_{31} & A_{32} & A_{33} 
\end{array}
\right],
\]
then $B$ is written as 
\[
B =\left[ 
\begin{array}{c|c|c|c}
A_{11} & A_{12} & A_{13} & A_{12} \\
\hline
A_{21} & {\boldsymbol A}_{22} & A_{23} & {\boldsymbol A}_{22} \\
\hline
A_{31} & A_{32} & A_{33} & A_{32} \\
\hline
A_{21} & 0 & 0 &  {\boldsymbol A}_{22}
\end{array}
\right].
\]

Apply the Laplace expansion 
with respect to
rows $I$  and  rows $\widetilde{I} \setminus I$,
we obtain
\begin{equation}   \label{eq:jac1}
\det B = \xi^I_I \xi^{I- J_1-J_2}_{I-J_1-J_2}.
\end{equation}
On the other hand, 
by 
the Laplace expansion with respect to
columns $I \setminus J_1$  and  columns $(\widetilde{I} \setminus I) \cup J_1$, 
we have
\begin{equation}   \label{eq:jac2}
 \det B =\sum_{\begin{subarray}{c} 
K_1,K_2 \subset I;\\
K_1 \cap (I-J_1-J_2) = \emptyset;\\
K_2 \cap (I-J_1-J_2) = \emptyset
\end{subarray}}  
\epsilon(K_1;K_2) 
\xi^{I-K_1}_{I-J_1} \xi^{I-K_2}_{I-J_2}.
\end{equation}
Thus we verify (\ref{eq:jacobi}).
\qed

\subsection{Vertex operators}
Introduce the vertex operators
$V_m(k;x,y)$ $(m \in {\mathbb  Z})$ 
defined by
(see \cite{T3})
\begin{equation}
V_m(k;x,y) = e^{m \xi(x -\widetilde{\partial}_{y},k)} e^{ -m \xi(\widetilde{\partial}_{x},k^{-1})  },
\end{equation}
where
$\widetilde{\partial}_{x}$ stands for 
$\left(\frac{\partial}{\partial x_1},\frac{1}{2} \frac{\partial}{\partial x_2},
\frac{1}{3} \frac{\partial}{\partial x_3}, \ldots\right)$
 and   
$\xi(x,k)=\sum_{n=1}^\infty x_n k^n$.
Define the differential operators 
$X_n$ and $Y_n$ $(n\in{\mathbb Z})$
by
\begin{equation}
\begin{array}{l}
 X(k) = \sum_{n \in {\mathbb Z} } X_n k^n=V_1(k;x,y) , \\
 Y(k) = \sum_{n \in {\mathbb Z} } Y_n k^{-n}=V_1(k^{-1};y,x).
\end{array}
\end{equation}
We have the following lemmas; see \cite{T3}.

\begin{lemma} 
\label{lem:uc}
The operators $X_n$ and $Y_n$ $(n \in {\mathbb Z})$
are raising operators for the universal characters 
in the sense that
\begin{equation}  \label{eq:uc}
S_{ [ \lambda,\mu ] } (x,y) 
= X_{ \lambda_1 } \cdots X_{\lambda_l} 
  Y_{\mu_1} \cdots Y_{\mu_{l'}} \cdot 1.                
\end{equation}
\end{lemma}

\begin{lemma}  
The following relations hold{\rm:}
\begin{equation}
\begin{array}{l}
X_mX_n + X_{n-1}X_{m+1} = 0, \\
Y_mY_n + Y_{n-1}Y_{m+1} = 0, \\
{}[X_m, Y_n]= 0,
\end{array}
\end{equation}
for $m,n \in {\mathbb Z}$.
In particular
$X_nX_{n+1}=Y_nY_{n+1}=0$.
\end{lemma}

\subsection{Proof of Theorem \ref{thm:Tmnuc}}

Introduce the Euler operator
\begin{equation}
E = 
\sum_{n=1}^\infty \left( 
n x_n \frac{\partial}{\partial x_n}
-n y_n \frac{\partial}{\partial y_n}
\right), 
\end{equation}
and operators 
$L^+$, $L^-$
given as follows:
\begin{eqnarray}
L^+ = \frac{x_1^2}{2}
+\sum_{n=1}^\infty \left( 
(n+2) x_{n+2} \frac{\partial}{\partial x_n}
-n y_n \frac{\partial}{\partial y_{n+2}}
\right)
-x_1 \frac{\partial}{\partial y_1}
-\left(-\kappa_\infty + \sum_i \theta_i \right)  
\frac{\partial}{\partial y_2},  \\
L^- = \frac{y_1^2}{2}
+\sum_{n=1}^\infty \left( 
(n+2) y_{n+2} \frac{\partial}{\partial y_n}
-n x_n \frac{\partial}{\partial x_{n+2}}
\right)
-y_1 \frac{\partial}{\partial x_1}
-\left(-\kappa_\infty + \sum_i \theta_i \right)  
\frac{\partial}{\partial x_2}.
\end{eqnarray}
Note that $E$, $L^+$, and $L^-$ are 
homogeneous operators of degrees
$0$, $2$, and $-2$, 
respectively.
Consider the change of the variables  
\begin{equation}
x_n=\frac{-\kappa_\infty+\sum_i \theta_i t_i^n}{n},
\quad 
y_n=\frac{-\kappa_\infty+\sum_i \theta_i t_i^{-n}}{n},
\end{equation}
and
\begin{equation}  \label{eq:tildeT}
\widetilde{T}_{m,n}(x,y) = (-1)^{- {v(v+1)}/{2}} 
             \prod_i t_i^{- {v(v+1)}/{2}} 
T_{m,n}(t),
\end{equation}
where
$u=|n-m-{1}/{2}|-{1}/{2}$,  
$v=|n+m-{1}/{2}|-{1}/{2}$.
Substitute this
into 
(\ref{subeq:Tmn}),
we have the 
recurrence relations
for $\widetilde{T}_{m,n}(x,y)$:
\begin{subequations}  \label{subeq:tilT}
\begin{eqnarray}
\lefteqn{-\widetilde{T}_{m+1,n}  \, \widetilde{T}_{m-1,n} }
\nonumber \\
&& 
= \left\{\left(L^- + E -\frac{y_1^2}{2} -2\right)
  \left(L^+-E-\frac{x_1^2}{2}\right) 
\log \widetilde{T}_{m,n}
-x_1 y_1 +(2 m )^2 \right\}
\widetilde{T}_{m,n}{}^2,
\qquad \qquad  
\label{eq:tilT1} \\
\lefteqn{- \widetilde{T}_{m,n+1}  \, \widetilde{T}_{m,n-1} }
\nonumber \\
&&  
= \left\{\left(L^- + E -\frac{y_1^2}{2} -2\right)
  \left(L^+-E-\frac{x_1^2}{2}\right) 
\log \widetilde{T}_{m,n}
-x_1 y_1 +(2 n-1 )^2 \right\}\widetilde{T}_{m,n}{}^2,
\label{eq:tilT2}
\end{eqnarray}
\end{subequations}
where the initial values are given by
\begin{equation}  \label{eq:tilT3}
\widetilde{T}_{0,0}= \widetilde{T}_{0,1} =1, \quad
\widetilde{T}_{1,0}= x_1, \quad
\widetilde{T}_{1,1}= y_1.
\end{equation}
Note that we have 
\begin{equation}  \label{eq:symtilT}
\widetilde{T}_{-m,n}(x,y) = \widetilde{T}_{m,1-n}(x,y) 
= \widetilde{T}_{m,n}(y,x),
\end{equation}
from (\ref{eq:symT}).

Theorem~\ref{thm:Tmnuc}
follows immediately  
from the
\begin{prop}  \label{prop:tilT}
Let
\begin{equation}  \label{eq:tilTS}
\widetilde{T}_{m,n}(x,y) = 
\prod_{j=1}^u(2j-1)!!
\prod_{k=1}^v(2k-1)!!  \,
S_{[\lambda,\mu]}(x,y),
\end{equation}
where 
$\lambda = (u,u-1, \ldots,2,1)$
and
$\mu=(v,v-1,\ldots,2,1)$,
then $\widetilde{T}_{m,n}(x,y)$ satisfies
{\rm (\ref{subeq:tilT})}
and
{\rm (\ref{eq:tilT3})}.
\end{prop}

We prepare some lemmas to verify Proposition~\ref{prop:tilT}. 

\begin{lemma}
The following commutation relations hold for 
$n \in {\mathbb Z}${\rm:}
\begin{eqnarray}
&&[X_n,L^+] = -\left(n+ \frac{3}{2}\right) X_{n+2}
+ 2\left(x_2-\frac{\partial}{\partial y_2} \right)X_n  , 
\label{eq:XL}
\\   
&&[Y_n,L^+] = \left(n-\frac{3}{2}-\kappa_\infty+\sum_i \theta_i\right)Y_{n-2}
-Y_n  \frac{\partial}{\partial y_2}, 
\label{eq:YL}
\\
&&
[X_n,x_2] = -\frac{1}{2} X_{n+2},  
\label{eq:Xx} \\
&&
[Y_n,x_2] = -\frac{1}{2} Y_{n-2}.
\label{eq:Yx}
\end{eqnarray}
\end{lemma}

\pf
Notice that
for any operators $A$ and $B$,
\[
e^A B e^{-A} = e^{{\rm ad}(A)} B 
= B +[A,B]+ \frac{1}{2!}[A,[A,B]]+\cdots,
\]
where 
${\rm ad}(A)(B)=[A,B]$.
We have
\[
[\xi(x -\widetilde{\partial}_{y},k),L^+] 
= - \sum_{m=1}^\infty 
\left\{  (m+2)x_{m+2} - \frac{\partial}{\partial y_{m+2}}
\right\}k^m, 
\]
so that
\begin{equation}  \label{eq:el1}
[e^{\xi(x -\widetilde{\partial}_{y},k)},L^+] 
= - \sum_{m=1}^\infty 
\left\{  (m+2)x_{m+2} - \frac{\partial}{\partial y_{m+2}}
\right\}k^m e^{\xi(x -\widetilde{\partial}_{y},k)}. 
\end{equation}
On the other hand,
we have
\begin{eqnarray*}
{}[-\xi(\widetilde{\partial}_{x},k^{-1}),L^+] 
&=& 
-\left(x_1 - \frac{\partial}{\partial y_1}\right)k^{-1} - \sum_{m=1}^\infty k^{-m-2} 
  \frac{\partial}{\partial x_m}, \\
{}[ -\xi(\widetilde{\partial}_{x},k^{-1})   ,[-\xi(\widetilde{\partial}_{x},k^{-1}),L^+] ]
&=&  k^{-2},
\end{eqnarray*}
then
\begin{equation}  \label{eq:el2}
[e^{-\xi(\widetilde{\partial}_{x},k^{-1})},L^+] 
= \left\{ 
-\left(x_1 - \frac{\partial}{\partial y_1}\right)k^{-1} 
+ \frac{k^{-2}}{2}
-\sum_{m=1}^\infty k^{-m-2} \frac{\partial}{\partial x_m}
\right\}
e^{-\xi(\widetilde{\partial}_{x},k^{-1})}.
\end{equation}
Noticing
\[
k^{-1}\frac{\partial}{\partial k} X(k) 
=\sum_{m=1}^\infty 
\left(  m x_{m} - \frac{\partial}{\partial y_{m}}  
\right)k^{m-2} X(k)
+ 
e^{\xi(x -\widetilde{\partial}_{y},k)}
\sum_{m=1}^\infty 
k^{-m-2} \frac{\partial}{\partial x_m}
e^{-\xi(\widetilde{\partial}_{x},k^{-1})},
\]
from (\ref{eq:el1}) and  (\ref{eq:el2}),
we obtain
\begin{eqnarray}
[X(k),L^+]&=&  e^{\xi(x -\widetilde{\partial}_{y},k)}
[e^{-\xi(\widetilde{\partial}_{x},k^{-1})},L^+] 
+
[e^{\xi(x -\widetilde{\partial}_{y},k)},L^+] 
e^{-\xi(\widetilde{\partial}_{x},k^{-1})} 
\nonumber \\
&=&
\left\{ 
-k^{-1} \frac{\partial}{\partial k}
+\frac{k^{-2}}{2} 
+2 \left(x_2 - \frac{1}{2}\frac{\partial}{\partial y_2}\right)
  \right\} X(k).
\end{eqnarray}
Take the coefficient of $k^n$,
we verify (\ref{eq:XL}).

We have
\begin{eqnarray*}
&&[\xi(y -\widetilde{\partial}_{x},k^{-1}),L^+]
= 
k^{-1}  \frac{\partial}{\partial y_1}
+\left(-\kappa_\infty +\sum_i \theta_i\right)k^{-2}
+\sum_{m=1}^\infty 
 \left(m y_m - \frac{\partial}{\partial x_m}\right)k^{-m-2},
\\
&&{} [\xi(y -\widetilde{\partial}_{x},k^{-1}), [\xi(y -\widetilde{\partial}_{x},k^{-1}),L^+]]
= -k^{-2},
\\
&&{}[-\xi(\widetilde{\partial}_{y},k),L^+]
=
 \sum_{m=1}^\infty k^m 
  \frac{\partial}{\partial y_{m+2}}, 
\end{eqnarray*}
so that 
\begin{eqnarray*}
{}[e^{\xi(y -\widetilde{\partial}_{x},k^{-1})},L^+]
&=& \left\{
 k^{-1}\frac{\partial}{\partial y_1} 
+ \left(-\kappa_\infty +\sum_i \theta_i-\frac{1}{2}
\right)k^{-2}
+ \sum_{m=1}^\infty 
 \left(m y_m - \frac{\partial}{\partial x_m}\right)k^{-m-2} 
\right\} ,
\\
{}[e^{-\xi(\widetilde{\partial}_{y},k)},L^+] 
&=&
\sum_{m=1}^\infty k^m \frac{\partial}{\partial y_{m+2}}
e^{-\xi(\widetilde{\partial}_{y},k)}.
\end{eqnarray*}
Thus we obtain
\begin{eqnarray}
[Y(k),L^+]&=&  e^{\xi(y -\widetilde{\partial}_{x},k^{-1})}
[e^{-\xi(\widetilde{\partial}_{y},k)},L^+] 
+
[e^{\xi(y -\widetilde{\partial}_{x},k^{-1})},L^+] 
e^{-\xi(\widetilde{\partial}_{y},k)} \nonumber \\
&=&
\left\{
-k^{-1} \frac{\partial}{\partial k}
+\left(-\kappa_\infty+\sum_i \theta_i +\frac{1}{2}\right)k^{-2}
\right\} Y(k) 
- Y(k) \frac{\partial}{\partial y_2},
\end{eqnarray}
whose coefficient of $k^{-n}$ yields (\ref{eq:YL}).

By
$[-\xi(\widetilde{\partial}_{x},k^{-1}),x_2] = -k^{-2}/2$,
we have
\[
[e^{-\xi(\widetilde{\partial}_{x},k^{-1})},x_2] 
= -\frac{k^{-2}}{2} 
e^{-\xi(\widetilde{\partial}_{x},k^{-1})},
\]
therefore
\begin{equation}
[X(k),x_2]= -\frac{k^{-2}}{2} X(k), \quad
[Y(k),x_2]= -\frac{k^{-2}}{2} Y(k).
\end{equation}
Take the coefficients of $k^n$ and $k^{-n}$,
we obtain (\ref{eq:Xx}) and (\ref{eq:Yx}) 
respectively. 
\qed

\begin{lemma}  \label{lem:LS}
For integers $u,v \geq 0$, 
the following formulae hold{\rm:}
\begin{eqnarray}
&&
L^+ S_{[u!,v!]}(x,y)
=(2u+1)S_{[(u+2,u-1,\ldots,1),v!]}(x,y)-(2u+1)x_2  S_{[u!,v!]}(x,y),
\label{eq:L+S}
\\
&&
L^- S_{[u!,v!]}(x,y)
=(2v+1)S_{[u!,(v+2,v-1,\ldots,1)]}(x,y)-(2v+1)y_2  S_{[u!,v!]}(x,y),
\label{eq:L-S}
\\ 
&&
L^+ S_{[u!,(v+2,v-1,\ldots,1)]}(x,y)
=(2u+1)S_{[(u+2,u-1,\ldots,1),(v+2,v-1,\ldots,1)]}(x,y)
\nonumber \\
&& \qquad \qquad \qquad  \qquad \qquad \qquad 
-(2u+1)x_2  S_{[u!,(v+2,v-1,\ldots,1) ]}(x,y) 
\nonumber \\
&& \qquad \qquad \qquad  \qquad \qquad \qquad 
-\left(v-u-\kappa_\infty+\sum_i \theta_i\right) S_{[u!,v!]}(x,y),
\label{eq:L+S-}
\\
&&
L^- S_{[(u+2,u-1,\ldots,1),v!]}(x,y)
=(2v+1)S_{[(u+2,u-1,\ldots,1),(v+2,v-1,\ldots,1)]}(x,y)
\nonumber \\
&& \qquad \qquad \qquad  \qquad \qquad \qquad 
-(2v+1)y_2  S_{[(u+2,u-1,\ldots,1),v! ]}(x,y) 
\nonumber \\
&& \qquad \qquad \qquad  \qquad \qquad \qquad 
-\left(u-v-\kappa_\infty+\sum_i \theta_i  \right) S_{[u!,v!]}(x,y).
\label{eq:L-S+}
\end{eqnarray}
Here 
$u!=(u,u-1, \ldots,2,1)$.
\end{lemma}

\pf
First we shall show that
\begin{equation}  \label{eq:L+Schur}
L^+ S_{[u!,\emptyset]}(x,y)=(2u+1)S_{[(u+2,u-1,\ldots,1),\emptyset]}(x,y)-(2u+1)x_2  S_{[u!,\emptyset]}(x,y),
\end{equation}
by induction.
Using
$S_{[\emptyset,\emptyset]}(x,y)=1$ and
$S_{[(2),\emptyset]}(x,y)=x_1^2/2 +x_2$,
it is easy to verify for $u=0$.
Assume that (\ref{eq:L+Schur}) 
is true for $u-1$.
Applying $X_u$,
we have
\begin{eqnarray*}
X_u L^+ S_{[(u-1)!,\emptyset]}(x,y) 
&=&  L^+ S_{[u!,\emptyset]}(x,y) +[X_u,L^+] S_{[(u-1)!,\emptyset]}(x,y) 
\nonumber \\
&=& (L^+ +2 x_2) S_{[u!,\emptyset]}(x,y)-
\left(u+\frac{3}{2} \right)
S_{[(u+2,u-1,\ldots,1),\emptyset]}(x,y),
\end{eqnarray*}
and
\begin{eqnarray*}
\lefteqn{X_u \left((2u-1)S_{[(u+1,u-2,\ldots,1),\emptyset]}(x,y)-(2u-1)x_2 S_{[(u-1)!,\emptyset]}(x,y)\right)}
\\
&& \qquad = - (2u-1)x_2 S_{[u!,\emptyset]}(x,y) + \frac{1}{2} (2u-1)  S_{[u!,\emptyset]}(x,y),
\end{eqnarray*}
by using the commutation relations (\ref{eq:XL})
and the property $X_k X_{k+1}=0$. 
Then, 
by the assumption, 
we have the desired equation 
(\ref{eq:L+Schur}) 
immediately.
Applying $Y_v Y_{v-1} \cdots Y_1$
to (\ref{eq:L+Schur}) 
we obtain (\ref{eq:L+S}).
Here we recall
the commutation relations (\ref{eq:YL}),  (\ref{eq:Yx}),
and $Y_k Y_{k+1}=0$.

Since $L^-$ is the same as $L^+$
except exchanging $x$ with $y$,
we verify (\ref{eq:L-S}) immediately.

Notice that
$S_{[u!,v!]}(x,y)$ does not depend on 
$y_{2n}$ $(n=1,2, \ldots)$.
Applying $Y_{v+3}$ to (\ref{eq:L+S}), 
we have
\[
Y_{v+3} L^+ S_{[u!,v!]}(x,y)=
L^+ S_{[u!,(v+3,v,\ldots,1)]}(x,y)
+\left(v+\frac{3}{2}-\kappa_\infty+\sum_i \theta_i\right)
S_{[u!,(v+1)!]}(x,y),
\]
and
\begin{eqnarray*}
&&Y_{v+3} \left((2u+1)S_{[(u+2,u-1,\ldots,1),v!]}(x,y)-(2u+1)x_2  S_{[u!,v!]}(x,y) \right)  \\
&&  \qquad = (2u+1)S_{[(u+2,u-1,\ldots,1),(v+3,v,\ldots,1)]}(x,y)
- (2u+1) x_2 S_{[u!,(v+3,v,\ldots,1) ]}(x,y) 
\\
&&  \qquad  \quad
+\left(u+\frac{1}{2}\right) S_{[u!,(v+1)!]}(x,y).
\end{eqnarray*}
Thus we verify (\ref{eq:L+S-}).
Similarly
(\ref{eq:L-S+}) also holds.
\qed
\\

\noindent  
{\it Proof of Proposition \ref{prop:tilT}.} 
For the sake of simplicity,
we use the following notations:
\begin{equation}
\begin{array}{l}
S=S_{[u!,v!]}(x,y), \\
S^+=S_{[(u+2,u-1,\ldots,1),v!]}(x,y), \\
S^-=S_{[u!,(v+2,v-1,\ldots,1) ]}(x,y), \\
S^{+-}=S_{[(u+2,u-1,\ldots,1),(v+2,v-1,\ldots,1)]}(x,y). 
\end{array}
\end{equation}
We have
\begin{eqnarray}
\lefteqn{
\left(\left(L^- + E -\frac{y_1^2}{2} -2\right)
  \left(L^+-E-\frac{x_1^2}{2}\right) 
\log S
\right)S^2}  \nonumber \\ 
&&\qquad  \qquad \qquad
= \left(L^- + E -\frac{y_1^2}{2}\right)
  \left(L^+-E-\frac{x_1^2}{2}\right)S \cdot S  
\nonumber \\
&& \qquad   \qquad  \qquad
\quad -  \left(L^- + E -\frac{y_1^2}{2}\right)S \cdot   
\left(L^+-E-\frac{x_1^2}{2}\right) S   
\nonumber \\ 
&& \qquad  \qquad  \qquad
\quad 
- 2 \left( L^+-E-\frac{x_1^2}{2}\right) S \cdot S.
\end{eqnarray}
Since
$S_{[\lambda,\mu]}(x,y)$ is a 
 weighted  homogeneous polynomial of 
degree $|\lambda|-|\mu|$,
the Euler operator $E$ acts on it as 
\begin{equation}
E S_{[\lambda,\mu]}(x,y) = (|\lambda|-|\mu|)S_{[\lambda,\mu]}(x,y).
\end{equation}
Then
by Lemma \ref{lem:LS}
we have
\begin{eqnarray}
\lefteqn{
\left(\left(L^- + E -\frac{y_1^2}{2} -2\right)
  \left(L^+-E-\frac{x_1^2}{2}\right) 
\log S  
-x_1 y_1
\right)S^2}   \nonumber \\
&\quad 
=
(2u+1)(2v+1)S^{+-} S - (2u+1)(2v+1)S^+ S^{-} 
-(u-v)^2 S^2.
\end{eqnarray}

Now
let us substitute  
(\ref{eq:tilTS}) 
into the recurrence relations 
(\ref{subeq:tilT}).
By virtue of (\ref{eq:symtilT}),
it is enough to consider the cases
(I) $n-m-1/2>0$, $n+m-1/2>0$;
and
(II) $n-m-1/2<0$, $n+m-1/2>0$. 

First we deal with the case
(I),
that is, 
$m=(v-u)/2$,
$n=(u+v+2)/2$.
Substitute (\ref{eq:tilTS}) into 
the both sides of 
(\ref{subeq:tilT}), 
we have
\begin{eqnarray*}
 {\rm LHS \ of \ (\ref{eq:tilT1})} &=&-
(2u+1)(2v+1)C_{u,v} 
S_{[(u+1)!,(v-1)!]} \cdot
S_{[(u-1)!,(v+1)!]}, 
\\
 {\rm RHS \ of \ (\ref{eq:tilT1}) } &=& (2u+1)(2v+1)C_{u,v}
(S^{+-} S - S^+ S^{-}), 
\end{eqnarray*}
and 
\begin{eqnarray*}
 {\rm LHS \ of \ (\ref{eq:tilT2})} &=&-
(2u+1)(2v+1)C_{u,v}
S_{[(u+1)!,(v+1)!]} \cdot S_{[(u-1)!,(v-1)!]}, 
\\
 {\rm RHS \ of \ (\ref{eq:tilT2})} &=& (2u+1)(2v+1)C_{u,v}
(S^{+-} S - S^+ S^{-}+S^2),
\end{eqnarray*}
respectively.
Here we put 
$C_{u,v}=\left(\prod_{j=1}^u(2j-1)!!\prod_{k=1}^v(2k-1)!!\right)^2$.
Thus it is sufficient to prove
\begin{eqnarray} 
-S_{[(u+1)!,(v-1)!]} \cdot
S_{[(u-1)!,(v+1)!]}
 &=& S^{+-} S - S^+ S^{-}, \label{eq:SS1}\\
-S_{[(u+1)!,(v+1)!]} \cdot S_{[(u-1)!,(v-1)!]}
&=& S^{+-} S - S^+ S^{-}+S^2.  \label{eq:SS2}
\end{eqnarray}
By using Lemma \ref{lem:pl} below,
we  immediately verify
(\ref{eq:SS1}) and (\ref{eq:SS2}).

The verification for the case (II)
is the same.
\qed

\begin{lemma}  \label{lem:pl}
The following formulae hold{\rm:}
\begin{eqnarray}  \label{eq:pl1}
&&S_{[(u+1)!,(v+1)!]} \cdot S_{[(u-1)!,(v-1)!]} 
-S_{[(u+1)!,(v-1)!]} \cdot S_{[(u-1)!,(v+1)!]} 
+{S_{[u!,v!]} }^2 = 0,  
\\
&&S_{[(u+1)!,(v-1)!]} \cdot S_{[(u-1)!,(v+1)!]} 
-S_{[u!,(v+2,v-1,\ldots,1)]} \cdot S_{[(u+2,u-1,\ldots,1),v!]}
\nonumber \\
&& \qquad \qquad \qquad \qquad \qquad \qquad \qquad 
+S_{[(u+2,u-1,\ldots,1),(v+2,v-1,\ldots,1)]} \cdot S_{[u!,v!]}  = 0.  
\label{eq:pl2}
\end{eqnarray}
\end{lemma}

\pf
Consider a $(u+v+2) \times (u+v+2)$ matrix
\begin{equation}
M= \left[
\begin{array}{cc|cccccccc|cc}
q_1 & \multicolumn{1}{c}{q_0}& 0 & 0 & \cdots & & && \cdots & \multicolumn{1}{c}{0}& 0& 0 \\
\cline{3-10}
    &q_2 &q_1 & & & & & & & & & \\
    &    &q_3 &q_2 & & & & & & & & \\
 & & & \ddots &\ddots & & & & & & & \\
 & & &        &q_v &q_{v-1} &&&&&& \\
\cline{3-10}
 & \multicolumn{1}{c}{} & &        &\cdots&q_{v+1} & q_v& \cdots&&\multicolumn{1}{c}{}&& \\
 &\multicolumn{1}{c}{} & &        &\cdots&p_u & p_{u+1}& \cdots&&\multicolumn{1}{c}{}&& \\
\cline{3-10}
 & & &        &      &    &p_{u-1} & p_u   &&&& \\
 & & &        &      &    &        &\ddots &\ddots&&& \\
 & & &        &      &    &        &       & p_2&p_3&& \\ &&&&&&&&& p_1&p_2 & \\
\cline{3-10}
0&\multicolumn{1}{c}{0}&0& \cdots &&&& \cdots &0&\multicolumn{1}{c}{0}& p_0&p_1 
\end{array}
 \right],
\end{equation}
so that
$D= \det M = S_{[(u+1)!,(v+1)!]}(x,y)$.
Denote by $D[i_1,i_2,\ldots;j_1,j_2,\ldots]$
its minor determinant removing rows $\{i_k\}$ and  columns $\{j_k\}$.
It is easy to see that
\begin{equation}
 \begin{array}{l}
 D \left[ 1,v+1,v+2,u+v+2  ; 1,2,u+v+1,u+v+2  \right]
= S_{[(u-1)!,(v-1)!]}(x,y), \\
 D[1,v+1;1,2] =S_{[(u+1)!,(v-1)!]}(x,y), \\
 D[v+2,u+v+2;u+v+1,u+v+2] =  S_{[(u-1)!,(v+1)!]}(x,y) ,\\
  D[1,v+2;1,2] = D[v+1,u+v+2;u+v+1,u+v+2] = S_{[u!,v!]}(x,y).
 \end{array}
\end{equation}
Applying Theorem \ref{thm:jacobi}, 
we have
\begin{eqnarray}
\lefteqn{D D[1,v+1,v+2,u+v+2;1,2,u+v+1,u+v+2]}  
\nonumber \\
&& 
= D[1,v+1;1,2]  
 D[v+2,u+v+2;u+v+1,u+v+2]  
\nonumber \\
&& \quad
- D[1,v+2;1,2]  
D[v+1,u+v+2;u+v+1,u+v+2],
\end{eqnarray}
which coincides with (\ref{eq:pl1}).

Take a $(u+v+2) \times (u+v)$ matrix
\begin{equation}
\widetilde{M}= \left[
\begin{array}{|cccccccc|}
\hline
q_1 & q_0 &0&\cdots &&&\cdots&0 \\
   & \ddots & \ddots&&&&& \\
   &    &q_{v-1} &q_{v-2}&&&& \\
\hline
\multicolumn{1}{c}{}& &\cdots& q_v&q_{v-1}&\cdots&&\multicolumn{1}{c}{} \\
\multicolumn{1}{c}{} &&\cdots&q_{v+2}&q_{v+1}&\cdots&&\multicolumn{1}{c}{} \\
\multicolumn{1}{c}{} &&\cdots&p_{u+1}&p_{u+2}&\cdots&&\multicolumn{1}{c}{} \\
\multicolumn{1}{c}{} &&\cdots&p_{u-1}&p_{u}&\cdots&&\multicolumn{1}{c}{} \\
\hline
&&&&p_{u-2}& p_{u-1}&&  \\
&&&& &\ddots&\ddots& \\
0&\cdots&&&\cdots&0&p_0&p_1 \\
\hline
\end{array}
 \right],
\end{equation}
then 
\begin{equation}
  \begin{array}{l}
D[v,v+1;\emptyset] =S_{[(u+1)!,(v-1)!]}(x,y), \quad
D[v+2,v+3;\emptyset] = S_{[(u-1)!,(v+1)!]}(x,y), \\ 
D[v,v+2;\emptyset] =S_{[u!,(v+2,v-1,\ldots,1)]}(x,y), \quad
D[v+1,v+3;\emptyset]  =S_{[(u+2,u-1,\ldots,1),v!]}(x,y), \\
D[v,v+3;\emptyset] = S_{[(u+2,u-1,\ldots,1),(v+2,v-1,\ldots,1)]}(x,y), \quad
D[v+1,v+2;\emptyset] =   S_{[u!,v!]}(x,y). 
  \end{array}
\end{equation}
By the Pl\"ucker relation, 
we have
\begin{eqnarray}
&&D[v,v+1;\emptyset] D[v+2,v+3;\emptyset]-
D[v,v+2;\emptyset] D[v+1,v+3;\emptyset]  
\nonumber \\
&&\qquad +D[v,v+3;\emptyset] D[v+1,v+2;\emptyset]=0,
\end{eqnarray}
which coincides with (\ref{eq:pl2}).
\qed
\\

\noindent
{\it Acknowledgement.}  
\small
The author wishes to thank 
Professor Kazuo Okamoto
for valuable discussions.
This work is partially supported by a fellowship of the Japan Society for the Promotion of Science (JSPS).


\end{document}